--------------------------------------------------------------------------
\documentstyle[eqsecnum,aps]{revtex}
\begin{document}

\title{Overtone spectra and intensities of tetrahedral molecules\\
in boson-realization models}

\author{Xi-Wen Hou }

\address{Institute of High Energy Physics, 
P.O. Box 918(4), Beijing 100039,\\
and Department of Physics, University of Three Gorges, 
Yichang 443000, The People's Republic of China }

\author{Mi Xie }

\address{ Department of Physics, Tianjin Normal University, 
Tianjin 300074, The People's Republic of China }

\author{Shi-Hai Dong}

\address{Institute of High Energy Physics, P.O. Box 918(4), 
Beijing 100039, The People's Republic of China }

\author{Zhong-Qi Ma}

\address{China Center for Advanced Science and Technology (World
Laboratory), P.O. Box 8730, Beijing 100080, \\
and Institute of High Energy Physics, 
P.O. Box 918(4), Beijing 100039, The People's Republic of China }

\maketitle

\vspace{50mm}
\begin{abstract}
The stretching and bending vibrational spectrum and
the intensities of infrared transitions in a tetrahedral molecule
are studied in two boson-realization models, where 
the interactions between stretching and bending vibrations 
are described by a quadratic cross term and by Fermi resonance terms,
called harmonically coupled and Fermi resonance boson-realization model,
respectively. The later is a development of our recent
model. As an example, the two models are applied to the 
overtone spectrum and the intensities of silicon tetrafluorde. 
Those models provide fits to the published experimental vibrational 
eigenvalues with standard deviations 1.956 cm$^{-1}$ and 
0.908 cm$^{-1}$, respectively. The intensities of infrared 
transitions of its complete vibrations are calculated in the 
two models, and results show a good agreement with the
observed data.

\end{abstract}

\newpage

\section {INTRODUCTION}

In recent years, algebraic methods have been introduced for 
a description of rotation-vibrational spectra of molecules. 
A U(4) algebraic model [1] was successfully used to explain 
the rotation-vibrational states of diatomic molecules.  
This model was developed for studying small molecules by
introducing a U(4) algebra for each bond [2,3].
It was also suggested to use a U$(k+1)$ model [4]
for the $k=3m-3$ rotational and vibrational degrees of freedom 
of $m$-atomic molecules. Those two models have an advantage that 
they can treat rotations and vibrations simultaneously, but they 
are quite complex for larger molecules. A U$(n)$ algebraic 
approach [5] was also used for a treatment of $n-1$ stretching 
vibrational degrees of freedom in polyatomic molecules. Iachello 
and Oss presented a SU(2) algebraic model based on isomorphism
between the one-dimensional Morse oscillator and the SU(2) 
algebra, which was particularly well suited for dealing with
the stretching vibrations of polyatomic molecules such as the 
octahedral and benzene-like systems [6]. The U(4) and SU(2) 
algebraic models were even modified by the corresponding quantum 
algebraic ones [7,8] for diatomic molecules.

Incorporating the bending modes in algebraic models for large
molecules was not achieved until 1993, when Iachello and Oss
proposed a U(2) algebraic model for describing P\"{o}schl-Teller 
oscillator that was well suitable for bending vibrations [9]. They 
extended this method to treat coupled bending modes [10]. Frank 
{\it et al.} presented a symmetry-adapted algebraic model [11-13], 
which described the stretching and bending vibrations in terms 
of U(2) algebra and made a clear connection between the algebraic 
approaches and the traditional methods in the configuration space.
In a different way, by making use of the bosonic operators for 
describing vibrations, Ma {\it et al.} have recently introduced an 
algebraic model of boson realization [14] for the complete 
vibrational modes, and obtained the satisfactory results for some 
molecules [15,16].

In this paper we will further investigate both the complete 
vibrations and the intensities of infrared transitions in 
a tetrahedral molecule in two boson-realization models. 
The first model is the model that we proposed recently [14], where the 
interactions between the stretch and the bend are harmonically 
coupled. Let us refer this model as a harmonically coupled 
boson-realization model (HCBM). The second model is called Fermi 
resonance boson-realization model (FRBM), where the interactions 
between two kinds of vibrations are described by Fermi resonances.
As an example of their applications, we study the vibrational 
spectrum of silicon tetrafluorde SiF$_4$ in two models. HCBM with 
seven parameters and FRBM with ten parameters provide fits to the 
observed values with the standard deviation 1.956 cm$^{-1}$ and 
0.908 cm$^{-1}$, respectively. Furthermore, we propose another
FRBM with only seven parameters, where the standard deviation is
still about half of that in HCBM with the same number of parameters. 
It shows that FRBM is more suitable for highly excited states 
in this molecule. In addition, intensities of infrared transitions of 
both the stretching and the bending vibrational spectrum in this 
molecule are calculated in the two models, and indicate a good 
agreement with the experimental values. 

The organization of this paper is as follows. Sec. II is devoted 
to construct the symmetrized bases and identify the spurious states. 
In Sec. III the vibrational Hamiltonian in the two models are 
introduced in terms of ten sets of boson operators and applied to 
the vibrational spectrum of silicon tetrafluoride SiF$_4$.
Its intensities of infrared transitions in the two models are presented 
in Sec. IV. Conclusion is made in Sec. V.

\section{SYMMETRIZED BASES AND SPURIOUS STATES}

In calculating spectra of polyatomic molecules one needs to construct 
a basis in which the Hamiltonian matrix is a block matrix. 
The symmetry adapted bases are widely used for this purpose. Halonen and 
Child [17] gave symmetrized local mode basis functions for stretching 
vibrations of symmetry molecules by a combination of promotion 
operators and Schmidt orthogonalization. Frank {\it et al.} [11-13] 
first constructed the symmetrized bases by projecting the one-phonon local 
functions, and then obtained the higher-phonon functions from the 
one-phonon symmetrized states by the Clebsch-Gordan coefficients. 
This method is quite complex when describing vibrations of large 
molecules and for high overtones. In this case, the symmetry adapted 
bases can be achieved by a new technique for constructing 
representations of the molecular symmetry point groups, that was 
recently called symmetrized boson representations [18]. This
new technique has merits that the basis vectors of those representations 
have a clear physical picture, and that their combinations are 
much simpler and general for multiple-phonon states. In this way 
we constructed the symmetrized bases for a tetrahedral molecule [14]. 
We hereby outline them for completeness. 

For a tetrahedral molecule XY$_{4}$ there are four stretching
oscillators and six bending oscillators. As in our previous paper [14],
let the atom X locate at the center O of the tetrahedron, 
and the four atoms Y at its vertices A, B, C, and D. 
The coordinate axes x, y, and z point from O to the centers 
of edges AC, AD, and AB, respectively. The four stretching 
oscillators OA, OB, OC, and OD, which are enumerated  one 
to four, are described by four equivalent bosonic operators 
$a^{\dagger}_{j}$ ($a_{j}$), $1\leq j \leq 4$. The six bending 
oscillators $\angle$ AOB, $\angle$ AOC, $\angle$ AOD, $\angle$ COD, 
$\angle$ DOB, and $\angle$ BOC, enumerated by five to ten, are 
described by six equivalent  bosonic operators $a^{\dagger}_{\mu}$ 
($a_{\mu}$), $5\leq \mu \leq 10$. Those ten sets of bosonic 
operators satisfy the well known algebraic relations. Hereafter, 
the indexes $j$, $\mu$, and $\alpha$ run from 1 to 4, 5 to 10, and 1 
to 10, respectively, and $n_{\alpha}$ denotes the phonon number on 
the $\alpha$th oscillator, and $n_s$ and $n_b$ the total phonon 
numbers on stretching and bending oscillators, respectively.

From the standard method of group theory, it is easy to reduce
the regular representation of the point group $T_{d}$. Applying
those combinations of group elements, that belong to irreducible 
representations of $T_{d}$, to the states $|n_{1}n_{2}n_{3}n_{4}\rangle$ 
for pure stretching vibrations and the states 
$|n_{5}n_{6}n_{7}n_{8}n_{9}n_{10}\rangle$ for pure bending vibrations, 
respectively, we generally obtain the symmetry adapted bases 
for the states with any phonon number. Then, the product of two bases 
can be combined into irreducible bases by the Clebsch-Gordan 
coefficients of $T_{d}$, denoted as $|n\rangle$ $\equiv$
$|n_{1}n_{2}n_{3}n_{4}n_{5}n_{6}n_{7}n_{8}n_{9}n_{10}\rangle$. 

For the fundamental bending vibrations ($n_{b}=1$) there are six
states belonging to three irreducible representation: $A_{1}$,
$E$, and $F_{2}$. As is well known [19], there are only five 
degrees of freedom for the bending vibrations ($E\oplus F_{2}$),
because there is a constraint between the six angles. The state 
belonging to the representation $A_{1}$ is called the fundamental 
spurious state:
\begin{equation}
\psi(A_{1},100000)=\displaystyle 6^{-1/2}~\sum_{\mu=5}^{10}
~|1_{\mu}\rangle ,
\end{equation}

\noindent
where $|1_{\mu}\rangle$ denotes the first excited state of the 
$\mu$th oscillator. This state introduces a spurious degree of 
freedom that should be removed.

Some methods for removing the spurious states were recently 
introduced. Iachello and Oss [9] placed the spurious states 
at the energies $\geq$ 10 times the energies of the physical states 
by projection operators. This method of removal is exact for harmonic   
bending vibrations and acquires a small error for anharmonic ones. 
Instead, Lemus and Frank [13] directly eliminated the spurious 
states from both the space and the Hamiltonian. They demanded the 
matrix elements of the Hamiltonian related to the fundamental 
spurious state $\psi(A_{1},100000)$ vanishing. However, it is 
impossible to demand all the matrix elements of the Hamiltonian 
that related to the spurious states vanishing. In Ref. [14] a 
criterion for identifying the spurious states was introduced: a state 
is a spurious state if it contains $\psi(A_{1},100000)$ as a factor. 
In addition, the bases of the physical states are chosen to be 
orthogonal with each other and with all the spurious states. In next 
section we will use this criterion to remove the spurious states 
from Hamiltonian and the symmetrized bases.

\section {HAMILTONIAN}

Studies of the vibrations in a tetrahedral molecule already exist 
in the literature. Its excited stretching vibrational states were 
explained in the local mode model [17] and a U(5) algebraic 
model [5]. A U(2) algebraic approach [13] was proposed for both 
stretching and bending modes, where the interactions between 
the stretch and the bend were neglected. In our previous papers [14,15] 
we studied the vibrational spectrum of methane in boson-realization 
model in terms of bosonic operators and q-deformed harmonic
oscillators, however, intensities of infrared transitions were not 
involved. We will further research both the vibrational energy levels 
and intensities of a tetrahedral molecules in the two different 
models, and apply it to the spectrum of SiF$_4$. 

\subsection {HARMONICALLY COUPLED BOSON-REALIZATION MODEL}

First of all, we outline the boson-realization model that we
recently introduced in [14], where the interactions
between the stretch and the bend are described by a quadratic 
cross term. We here call it as the harmonically coupled 
boson-realization model (HCBM). Then, we calculate the energy 
levels of SiF$_4$ in HCBM.  

For simplicity we assume that all oscillators are the Morse ones 
with two parameters $\omega$ and $x$, so that the energies of 
those oscillators can be expressed in the operator form:
$$E_s(n_j)=n_j\left\{{\omega}_s-x_s(n_j+1)\right\}, ~~~~
1\leq j \leq 4, $$
\begin{equation}
E_b(n_{\mu})=n_{\mu}\left\{{\omega}_b-
x_b(n_{\mu}+1)\right\},~~~~5\leq \mu \leq 10,
\end{equation}

\noindent
where the null energy has been removed.

The vibrational Hamiltonian $H$ of the molecule XY$_4$ is
$T_{d}$ invariant and is assumed to preserve the total number of 
quanta $n=n_{s}+n_{b}$. Since $a^{\dagger}_{j}$ ($a_{j}$) and 
$a^{\dagger}_{\mu}$ ($a_{\mu}$) are the tensor operators belonging to 
$A_{1}\oplus F_{2}$ and $A_{1}\oplus E \oplus F_{2}$, respectively, 
and $\sum a_{\mu}$ (or $\sum a_{\mu}^{\dagger}$) only annihilates 
(or creates) the spurious states, we can express Hamiltonian $H$ 
as the $T_{d}$ invariant combinations of the products of one 
creation operator and one annihilation operator [15]: 
$$H=\displaystyle \sum_{j=1}^{4}~E_{s}(a^{\dagger}_{j}a_{j})
+ \displaystyle \sum_{\mu=5}^{10}~E_{b}(a^{\dagger}_{\mu}a_{\mu})
+ \lambda_{s}\displaystyle \sum_{i\neq j}~a^{\dagger}_{i}a_{j} $$
$$~~~~+~\lambda_{b}\displaystyle \sum_{\mu=5}^{7}~\left(
a^{\dagger}_{\mu}a_{\mu+3} +{\rm H.c.}\right)
~+~\lambda_{sb}\left\{\displaystyle 
a^{\dagger}_{1}\sum_{\mu =5}^{7} \left(a_{\mu}-a_{\mu +3}\right) \right.$$
$$~~~~+~ a^{\dagger}_{2} \left(a_{5}-\displaystyle \sum_{\mu =6}^{8} a_{\mu}
+a_{9}+a_{10}\right) 
~~~~+a^{\dagger}_{3}\displaystyle \sum_{\mu =3}^{5}
 \left(a_{2\mu}-a_{2\mu -1}\right)  $$
\begin{equation}
~~~~+~\left.a^{\dagger}_{4}\left(-a_{5}-a_{6}+
\displaystyle \sum_{\mu =7}^{9} a_{\mu}
-a_{10}\right)+{\rm H.c.}\right\},
\end{equation}

\noindent
where $\lambda_s$ and $\lambda_{b}$ are the coupling strength 
among stretching modes and bending ones, respectively. The term 
with $\lambda_{sb}$ describes the interaction between stretching and
bending vibrations, which was neglected in Ref. [13] for simplicity.
The Hamiltonian $H$ contains seven parameters: $\omega_{s}$, $x_{s}$, 
$\omega_{b}$, $x_{b}$, $\lambda_{s}$, $\lambda_{b}$, 
and $\lambda_{sb}$. 

Following the criterion in Sec. II for eliminating the spurious 
states from the bases and straight calculating, we find that for
$n=1$ and representation $A_{1}$, there is one physical state, 
for $n=1$ and $E$ there is only one set of physical states, 
for n=1 and $F_{2}$ there are two sets of physical states, for 
n=2 and $F_{2}$ there are seven sets of physical states, 
and for n=3 and $F_{2}$, there are 25 sets of physical states, 
and for n=4 and $F_2$, there are 69 sets of physical states. 
Those combinations for $n=1$, 2 were given in Ref. [14], and 
others can be obtained from us upon request.

In those bases of the physical states the Hamiltonian becomes
a symmetric block matrix. The energies of physical states can
be calculated provided that the parameters of the Hamiltonian 
are known. A least-square fitting is adopted to get the parameters 
from the observed data. The observed energy levels from Ref. [20] 
and the corresponding calculated values are given in Table I, where 
the standard deviation (SD) in this fit (Fit A) is 1.956 cm$^{-1}$. 
The seven parameters obtained are given in Table II. In terms of 
those parameters one can calculate the other energy levels. Since 
the infrared experimental dipole transitions energies correspond to 
$F_2$, we hereby only list the calculation results for other $F_2$ 
states in Table III.

\begin{center}

\fbox{Table I}

\vspace{3mm}
\fbox{Table II}

\vspace{3mm}
\fbox{Table III}

\end{center}

From Table II we see that the ratio $|w_s/x_s|$ of the 
stretching modes is larger than that of the bending ones.
It means that, contrary to the case in methane [14], the 
anharmonicity of bending modes in $SiF_{4}$ is larger than 
that of the stretching ones. At least, the anharmonicity
of the vibrations depends upon the molecules discussed.

\subsection{FERMI RESONANCE BOSON-REALIZATION MODEL}
 
With the development and refinement of experimental techniques in 
high-resolution spectroscopy, measurements of high excited 
vibrational spectra for molecules become available [21]. One of 
important characters in highly excited states is appearance
of anharmonic resonances. Fermi resonance (FR), one of these resonances, 
is taken into account for description of molecular vibrations. 
In the normal mode model, FR terms were treated as perturbative 
corrections [22], while they were described by the nondiagonal 
matrix elements of Majorana operators in U(4) algebraic model [23]. 
Recently, simple Fermi resonance-local mode models for bent 
triatomic molecules [24] and pyramidal XY$_3$ molecules [25] have 
been constructed by Halonen {\it et al.}, where FR terms are expressed 
in terms of curvilinear internal valence coordinates. In addition, 
FR can be of central importance for intramolecular vibrational 
redistribution and kinetics [26]. Therefore, it is necessary for us 
to consider FR in the boson-realization model. We present an extended 
model, called Fermi resonance boson-realization model (FRBM),
and it is pleasure to see that FRBM provides smaller standard deviation
than HCBM in the energy level fits of SiF$_4$.

We introduce another vibrational Hamiltonian for a tetrahedral 
molecule, where the interactions 
between stretching and bending vibrations are described by the 
$T_{d}$ symmetric FR terms that couple one creation (or, 
respectively, annihilation) operator of stretching vibrations 
with two annihilation (or, respectively, creation) operators 
of bending ones. Although $a_{\mu}a_{\nu}$ belongs to 
$3A_{1}\oplus 3E\oplus 3F_{2}\oplus F_{1}$, we find only four 
independent $T_{d}$ invariant combinations related to 
physical states:

i) $A_{1}\otimes \left(E \otimes E\right)_{A_{1}}$,
\begin{equation}
H_{1}~=~\displaystyle \left(\sum_{j=1}^{4}~a_{j}^{\dagger}\right) 
\left(\sum_{\mu=5}^{10}~a_{\mu}^{2}- \sum_{\mu< \nu=6}^{10}a_{\mu}a_{\nu}
+ 3\sum_{\mu=5}^{7}~a_{\mu}a_{\mu+3}
\right)~+~{\rm H.c.},
\end{equation}

ii) $A_{1}\otimes \left(F_{2} \otimes F_{2}\right)_{A_{1}}$,
\begin{equation}
H_{2}~=~\displaystyle \left(\sum_{j=1}^{4}~a_{j}^{\dagger}\right) 
\left(\sum_{\mu=5}^{10}~a_{\mu}^{2}~-~2\sum_{\mu=5}^{7}~
a_{\mu}a_{\mu+3}\right)~+~H.c.,
\end{equation}

iii) $F_{2}\otimes \left(E \otimes F_{2}\right)_{F_{2}}$,
$$H_{3}~=~\left(a_{1}^{\dagger}-a_{2}^{\dagger}+a_{3}^{\dagger}
-a_{4}^{\dagger}\right)\left(a_{6}-a_{9}\right)
\left(3a_{6}+3a_{9}-\displaystyle \sum_{\mu=5}^{10}~a_{\mu}\right)$$
$$~~~~+~\left(a_{1}^{\dagger}-a_{2}^{\dagger}-a_{3}^{\dagger}
+a_{4}^{\dagger}\right)\left(a_{7}-a_{10}\right)
\left(3a_{7}+3a_{10}-\displaystyle \sum_{\mu=5}^{10}~a_{\mu}\right)$$
\begin{equation}
~~~~+~\left(a_{1}^{\dagger}+a_{2}^{\dagger}-a_{3}^{\dagger}
-a_{4}^{\dagger}\right)\left(a_{5}-a_{8}\right)
\left(3a_{5}+3a_{8}-\displaystyle \sum_{\mu=5}^{10}~a_{\mu}\right)+H.c., 
\end{equation}

iv) $F_{2}\otimes \left(F_{2} \otimes F_{2}\right)_{F_{2}}$,
$$H_{4}~=~\left(a_{1}^{\dagger}-a_{2}^{\dagger}+a_{3}^{\dagger}
-a_{4}^{\dagger}\right)\left(a_{5}-a_{8}\right) \left(a_{7}-a_{10}
\right) $$
$$~~~~+~\left(a_{1}^{\dagger}-a_{2}^{\dagger}-a_{3}^{\dagger}
+a_{4}^{\dagger}\right)\left(a_{5}-a_{8}\right) \left(a_{6}-a_{9}
\right) $$
\begin{equation}
~~~~+~\left(a_{1}^{\dagger}+a_{2}^{\dagger}-a_{3}^{\dagger}
-a_{4}^{\dagger}\right)\left(a_{6}-a_{9}\right) \left(a_{7}-a_{10}
\right)  +{\rm H.c.}.
\end{equation}

Now, we obtain the following $T_{d}$ invariant Hamiltonian with
ten parameters:
$$H=\displaystyle \sum_{j=1}^{4} ~E_{s}(a^{\dagger}_{j}a_{j})
+ \displaystyle \sum_{\mu=5}^{10} ~E_{b}(a^{\dagger}_{\mu}a_{\mu})
+ \lambda_{s}\displaystyle \sum_{i\neq j=5}^{10} a^{\dagger}_{i}a_{j}
+ \lambda_{b}\displaystyle \sum_{\mu=5}^{7}\left(
a^{\dagger}_{\mu}a_{\mu+3}+a^{\dagger}_{\mu+3}a_{\mu}\right) $$
\begin{equation}
~~~~+~\lambda_{f1}~H_{1}~+~\lambda_{f2}~H_{2}
~+~\lambda_{f3}~H_{3}~+~\lambda_{f4}~H_{4},
\end{equation}

\noindent
where the Hamiltonian preserve the quantum number $N=2n_s+n_b$.

The symmetrized bases are also used in the calculation. 
Using the same method for removing the spurious states 
as described in Sec. II, we find that there are 
1 set of states with $N=1$ for representation $E$ (no 
spurious state), and 3 states with N=2 for $A_{1}$ (removing 
1 spurious state), and for the representation $F_{2}$ there 
are 1 set of states with $N=1$ (no spurious state), 3 sets of 
states with $N=2$ (removing 1 set of spurious state), 8 sets of
states with $N=3$ (removing 4 sets of spurious states), 20 
sets of states with $N=4$ (removing 12 sets of spurious states), 
43 sets of states with $N=5$ (removing 32 sets of spurious states), 
and 90 sets of states with $N=6$ (removing 75 sets of spurious 
states). The physical states and the Hamiltonian matrices in 
those physical states are evaluated with the help of the 
computer algebra program MATHEMATICA [27]. 
    
For comparison, we also list the ten parameters determined by
fitting the same 16 experimental data (Fit B) in Table II and 
the calculated results in Table I, respectively. The standard 
deviation in this fit is 0.908 cm$^{-1}$. Other calculation
results for the spectrum only for $F_2$ are given in Table IV.

\begin{center}

\fbox{Table IV}

\end{center}

From Table II we see that the parameters $\lambda_{f2}$, 
$\lambda_{f3}$ and $\lambda_{f4}$ are quite small. It provides
us a possibility to set those three parameters vanishing. 
In this way another fitting (Fit C) is obtained with the 
standard deviation 0.989 cm$^{-1}$, where it contains the 
same number of parameters as in Fit A. 
This standard deviation is about one half of that in Fit A. 
The seven parameters and the calculated values obtained in 
Fit C are given in the corresponding tables. 

It is interesting to compare the obtained parameters in the three fits.
From Table II, when FR terms (Fit B) is replaced
with the harmonical coupling (Fit A) between the stretch and the bend,
the biggest change of the parameters is that of the anharmonic 
constant $x_s$ in Morse oscillator of the stretch,
while parameter $\lambda_b$ for the interactions between the bending
vibrations gets the least change. When the three small FR parameters
are set to be zero (Fit C), other parameters change very little 
in comparison with those in Fit B. 
Fit C provides less standard deviation than Fit A
with the same number of parameters. To our knowledge, it may 
be the model with the least parameters that well fits the observed
vibrational spectrum of SiF$_4$. In principle, the physical meaning 
of those obtained parameters can be explained in terms of 
internal coordinate.
 
It is worth mentioning that McDowell {\it et al.} [20] described 
the same vibrational spectrum of SiF$_4$ by the normal mode model with 
more parameters. This model provides the normal mode labels for 
small amplitude polyatomic molecular vibrational spectra. However, 
it becomes less appropriate at higher levels of vibrational 
excitation. In particular, the assignment of sets of normal mode 
quantum numbers to give spectral absorption features becomes 
inherently ambiguous [28]. This model does not provide explicitly
wave functions so that some physical properties such as transition 
intensities are hard to be calculated. 

\section{ INTENSITIES OF INFRARED TRANSITION}

Having completed calculations for the vibrational energy levels 
of SiF$_4$, we are now able to compute the intensities of infrared 
transition for all active modes. This information 
can be used to check the assignments and in the study of 
intramolecular energy relaxation in SiF$_4$. For stretching 
vibrations of tetrahedral molecules, Leroy {\it et al.} [29] 
recently constructed an electric dipole 
moment operator through unitary algebra and point group symmetry. 
However, their dipole function is not feasible for treating other 
vibrational modes. Intensities of infrared and Raman
transition of stretching modes in octahedral molecules were 
analyzed by Chen {\it et al.} [30] in terms of U(2) algebra. 
In this section, intensities of infrared transition for all active 
modes of SiF$_4$ will be computed in the two approaches presented 
in Sec. III. 

The absolute absorption intensities from state $n'$ to $n$ 
in the infrared active mode $F_2$ are given by
$$I_{nn'}=\nu_{nn'}P_{nn'},   $$
\begin{equation}
P_{nn'}=|\langle n|\hat{T}_x|n'\rangle|^{2}+
|\langle n|\hat{T}_y|n'\rangle |^{2}+
|\langle n|\hat{T}_z|n'\rangle |^{2},
\end{equation}

\noindent
where $\nu_{nn'}$ is the frequency of the observed transition,
and $\hat{T}_x$, $\hat{T}_y$, and $\hat{T}_z$ correspond to the three
components of the infrared transition operator $\hat{T}$. All other
constants are absorbed in the normalization of the operator $\hat{T}$.
The three components of $\hat{T}$ are
$$\hat{T}_x~=~\gamma_s~(\hat{t}_1-\hat{t}_2+\hat{t}_3-\hat{t}_4)
            ~+~\gamma_b~(\hat{t}_6-\hat{t}_9)
~+~\gamma_{sb}~(\hat{t}_1+\hat{t}_2+\hat{t}_3+\hat{t}_4)(\hat{t}_6
        -\hat{t}_9),$$
$$\hat{T}_y~=~\gamma_s~(\hat{t}_1-\hat{t}_2-\hat{t}_3+\hat{t}_4)
            ~+~\gamma_b~(\hat{t}_7-\hat{t}_{10})
~+~\gamma_{sb}~(\hat{t}_1+\hat{t}_2+\hat{t}_3+\hat{t}_4)(\hat{t}_7
      -\hat{t}_{10}),$$
\begin{equation}
\hat{T}_z~=~\gamma_s~(\hat{t}_1+\hat{t}_2-\hat{t}_3-\hat{t}_4)
            ~+~\gamma_b~(\hat{t}_5-\hat{t}_8)
~+~\gamma_{sb}~(\hat{t}_1+\hat{t}_2+\hat{t}_3+\hat{t}_4)(\hat{t}_5
     -\hat{t}_8),
\end{equation}

\noindent
where $\gamma_s$, $\gamma_b$, and $\gamma_{sb}$ are parameters,
and $\hat{t}_{\alpha}$ is the local operator on the $\alpha$th bond.
The term with $\gamma_{sb}$ is the higher order contribution of $\hat{T}$,
which is necessary for describing both the stretching and the bending
active modes in SiF$_4$. Following [9] we take the matrix 
elements of $\hat{t}_{\alpha}$ as follows:
$$\langle n|\hat{t}_{j}|n'\rangle~=~exp(-\eta_{j}|n_{j}
-n'_{j}|),~~~~1\leq j \leq 4, $$
\begin{equation}
\langle n|\hat{t}_{\mu}|n'\rangle~=~exp(-\eta_{\mu}|n_{\mu}
-n'_{\mu}|),~~~~5\leq \mu \leq 10,
\end{equation}

\noindent
where the coefficients $\eta_j$ should be equal for the equivalent bonds
$\eta_j$$\equiv$$\eta_s$, and $\eta_{\mu}$$\equiv$$\eta_b$.

Since calculations are done in the symmetrized bases it is 
sufficient to consider only the $z$ component, $\hat{T_z}$. 
All others can be obtained by making use of the Wigner-Eckart 
theorem. The relative intensities calculated in Fit A and Fit B
are given in Table I, where they are compared with experiment.
The corresponding parameters and the standard deviations (SD) are 
listed in Table V. In the two fits, the obtained parameters in 
the operator of infrared transition have a little difference, but
the calculated intensities for a few energy levels are quite
different. This is owing to the different wave functions and
the same transition operator in the two models.

\begin{center}

\fbox{Table V}

\end{center}

In Table I most of calculated intensities for the two models
are in good agreement with the experimental values, but a few 
are not. Those differences may come from two sources. The observed 
intensities are only approximately accurate [20], and the other 
higher order contributions to the operator $\hat{T}$ are neglected.
The more accurate experimental data are needed to improve the models. 

In order to complete the spectroscopic study of vibrational overtone
of SiF$_4$, we should also calculate the intensities of Raman transition 
in the two models. Unfortunately, We have to postpone this calculation 
due to lack of the observed values.

\section{CONCLUSION}

For studying the stretching and bending spectrum of a tetrahedral 
molecule, we have presented a harmonically coupled boson-realization 
model (HCBM) and a Fermi resonance boson-realization model (FRBM), 
where the coupling between the stretch and the bend is described 
by a quadratic cross term and by Fermi resonance terms, respectively.
The two models have been applied to 
the complete vibrations of silicon tetrafluoride SiF$_4$. HCBM with 
seven parameters and FRBM with ten parameters provide fits to the 
published experimental vibrational eigenvalues with standard 
deviations 1.956 cm$^{-1}$ and 0.908 cm$^{-1}$, respectively. 
This is based on our new method for constructing symmetrized
bases [14] and for removing both the spurious states in the
wavefunction space and the spurious components in Hamiltonian [15].
This method is particularly useful for highly excited states
in large molecules. In another FRBM, we decrease the number of parameters
and obtain the standard deviation 0.989 cm$^{-1}$ 
that is about half of that in HCBM with the same number of parameters. 
To our knowledge, FRBM may be the 
model for a good description of vibrational spectrum of SiF$_4$ 
with the least parameters. We believe that FRBM will be better suitable 
for describing highly excited vibrations in molecules.

The intensities of infrared transitions of the complete vibrations in 
this molecule have been calculated in those two models. The model
transition operator with five parameters well reproduces the 
observed data. The more satisfactory results can be obtained if
more accurate experimental values are available.

Finally, we remark that our models can be extended in several ways. 
The Fermi resonances may be taken into account as perturbative terms 
in HCBM, or the harmonically coupled term appears in 
FRBM as a perturbative one. Other anharmonic resonances
such as Darling-Dennison resonances can be included in the models
by adding higher-order terms of bosonic operators [16].
The rotational degrees of freedom can be incorporated by coupling 
the vibrational wave functions to rotational states carrying 
the appropriate point symmetries [31]. Our algebraic Hamiltonian 
can be written down in the coset space representation to study the 
dynamics of stationary eigenstates and inter-mode energy 
transfer [32]. Work on those subjects is in progress.

\acknowledgments 
The authors would like to thank Prof. Jin-Quan Chen and 
Dr. Jia-Lun Ping for useful discussions. This work was supported by 
the National Natural Science Foundation of China and Grant No. 
LWTZ-1298 of the Chinese Academy of Sciences.

\newpage
\begin{center}
{\small Table I. Observed and calculated energy levels and 
relative intensities for SiF$_4$}

\vspace{3mm}
\begin{tabular}{ccc|ccc|ccc|c}
\hline
\hline
   & Obs.[20] &    &   & Fit A &   &    & Fit B  &  & Fit C  \\ 
\hline
$~\Gamma~$&$~E$ (cm$^{-1}~$)&Intensity&$~n~$&$~E$ (cm$^{-1}~$)&Intensity
&$~N~$&$~E$ (cm$^{-1}~$) &Intensity & E($cm^{-1}$)\\ \hline
$E$    & 264.2   &      &1&262.593 &     &1& 262.616 & &263.043 \\
$F_2$& 388.4448& 500  &1&387.029 &499.966 &1& 388.112 &500.086
&387.552 \\
$F_2$& 776.3   & 0.9  &2&776.028 &0.968   &2& 776.196 &0.413
&775.104 \\
$A_1$&800.8    &      &1&803.333 &        &2& 800.195 & &799.847\\
$F_2$&1031.3968& 5000 &1&1031.338&4999.989&2& 1030.951& 5000.031
&1030.818\\
$E$    &1064.2   &      &2&1065.926&        &3& 1064.486& &1064.411\\
$F_2$&1164.2   & 1.4  &3&1164.348&4$\times10^{-4}$&3&1164.939&4.117
&1165.115\\
$F_2$&1189.7   & 40   &2&1190.361&40.073  &3& 1190.236&39.631
&1190.531\\
$F_2$&1294.05  & 2.4  &2&1293.938&1$\times10^{-4}$&3&1294.308&0.009
&1293.861\\
$F_2$&1418.75  & 0.1  &2&1418.367&2$\times10^{-4}$&3& 1418.989&
1$\times10^{-5}$ &1418.370\\
$F_2$&1804.5   & 0.7  &3&1805.405&2$\times10^{-6}$&4&1804.941&0.002
&1805.922\\
$F_2$&1828.17  & 7    &2&1828.098&3.820   &4& 1828.219 &3.909&1828.083\\
$F_2$&2059.1   & 1.2  &2&2056.461&3.653   &4& 2059.016 &4.193&2059.098\\
$F_2$&2602.55  & 0.007&4&2602.163&2$\times10^{-9}$&6& 2601.910 &
2$\times10^{-6}$& 2602.273\\ 
$F_2$&2623.8   & 0.015&3&2621.531&0.002   &6& 2624.028 & 0.007&2624.067\\ 
$F_2$&3068.5   & 0.015&3&3071.082&0.004   &6& 3068.276 & 0.005&3068.251\\     
\hline
\hline
\end{tabular}
\end{center}

\vspace{10mm}
\begin{center}
{\small Table II. Parameters in the Hamiltonian obtained by the least 
square fitting $\left(cm^{-1}\right)$}

\vspace{3mm}
\begin{tabular}{c|c|c|c|c}
\hline
\hline
&stretching&bending&interactions & \\ \cline{2-4}
&$\omega_{s}~~~~~~~x_{s}~~~~~~~\lambda_{s}$
&$\omega_{b}~~~~~~~x_{b}~~~~~~~\lambda_{b}$ 
&$\lambda_{sb}~(\lambda_{fk},~k=1,4)$ & SD\\
\hline
Fit A &987.106~~6.395~~~-56.995 & 135.205~~~-94.810~~-62.232
& 1.486 & 1.956 \\
Fit B &978.898~~2.930~~~-57.661 & 133.656~~~-95.854~~-62.748
& 1.274~~0.431~~0.772~~0.635 & 0.908 \\
Fit C &978.043~~2.567~~~-57.909 & 132.979~~~-96.159~~-62.255
& $1.167^{*}$ & 0.989 \\
\hline
\hline
\end{tabular}
\end{center}

\noindent
{\small ~~~~~~~ $^{*}~\lambda_{f2}=\lambda_{f3}=\lambda_{f4}=0$.}

\newpage
\begin{center}
{\small Table III. The calculated energy levels in Fit A for other 
$F_{2}$ states without observed data (in cm$^{-1}$)}

\vspace{3mm}
\begin{tabular}{rrrrrrrrr} 
\hline
\hline
 n=2 & 774.058 &&&&&&& \\ 
 n=3 &3088.993  &2846.495  &2449.719
   &2443.500  &2440.189  &2319.058  &2215.126  &2090.697  \\
   &1990.429  &1920.150  &1909.535  &1807.367  
   &1680.964  &1599.793  &1579.359  &1578.813 \\
   &1577.390  &1489.258  &1275.125  &1099.854  &944.679 &&& \\
n=4  &4097.150  &4082.620  &3873.222  &3853.713  &3633.127
    &3476.023  &3468.264  &3462.928 \\
 &3458.116  &3411.629  
    &3351.604  &3343.735  &3333.687  &3240.262  &3233.533  
    &3229.815  \\
&3109.090  &3008.558  &2945.283  &2934.648  
    &2884.130  &2838.738  &2836.739  &2832.567 \\
 &2830.527  
    &2829.199  &2827.224  &2787.228  &2716.908  &2712.270  
    &2706.295  &2706.110 \\
 &2624.924  &2604.126  &2603.943  
    &2520.595  &2477.722  &2444.796  &2432.932  
    &2396.553  \\
&2379.427  &2377.458  &2375.573  &2306.477  
    &2302.174  &2292.589  &2257.637  &2195.687 \\
 &2192.434  
    &2168.875  &2164.824  &2131.252  &2086.635  &2078.458  
    &1976.028  &1967.680 \\
 &1941.916  &1910.947  &1903.186  
    &1896.622  &1845.473  &1748.012  &1663.632  &1651.223 \\ 
    &1634.084  &1581.540  &1448.531  &1385.549 &&&&\\
\hline
\hline
\end{tabular}
\end{center}

\newpage
\vspace{4mm}
\begin{center}
{\small Table IV. The calculated energy levels in Fit B 
for other $F_{2}$ states without observed data (in cm$^{-1}$)}

\vspace{3mm}
\begin{tabular}{rrrrrrrrr} 
\hline
\hline
N=2 & 778.309 &&&&&&&  \\
N=3 &1496.547  &1279.868  &1102.618 &945.771 &&&& \\  
N=4 &2177.309     &1925.826     &1918.702   &1911.620     
&1855.411  &1809.079  &1682.927     &1669.183    \\
 &1655.980  &1644.807     &1600.570     &1588.721     &1578.839     
 &1575.762     &1569.344     &1451.455    \\
 &1388.306  &&&&&&& \\    
N=5    
   &3092.260    &2722.565    &2649.070    &2592.144    &2526.968   
   &2462.997    &2455.394    &2447.251  \\
  &2436.558 &2430.680   
   &2421.387    &2322.520    &2318.279    &2310.335    &2305.182   
   &2296.747  \\
  &2272.404  
  &2257.953    &2228.585    &2218.677   
   &2216.229    &2195.354    &2193.315    &2173.913 \\
 &2157.729 
   &2132.698    &2102.615    &2092.409    &2089.835    &2080.143   
   &2073.023    &2039.092  \\
  &1990.464   
 &1976.180    &1963.068   
   &1956.843    &1938.327    &1906.334    &1897.406    &1848.606\\   
   &1824.641   &1748.271    &1655.093   &&&&& \\
 N=6  
    &4078.063     &3543.827     &3517.840     &3426.476  &3261.511     
    &3258.607     &3217.764     &3207.488   \\
  &3203.984     &3090.515     
    &3082.210     &3046.639     &2992.877     &2977.422     
    &2955.982     &2952.235  \\
   &2946.590     &2941.518     &2939.587     
    &2928.298     &2903.491     &2901.129     &2894.450  &2886.484 \\    
    &2853.321     &2846.446     &2841.328     &2838.193  &2835.863     
    &2835.647     &2832.926     &2831.058  \\
   &2828.552     &2806.848     
    &2793.846     &2789.674     &2760.970     &2742.810  &2741.421     
    &2725.766   \\
  &2719.000     &2715.320     &2714.351     &2708.920     
    &2701.105     &2698.761     &2693.801     &2685.015  \\
   &2678.378     
    &2672.708     &2660.143     &2655.004     &2647.133  &2638.670     
    &2630.250     &2615.894    \\
 &2607.824     
    &2598.092     &2585.833     &2582.730     &2555.568  &2535.720     
    &2505.022     &2487.403   \\
  &2480.529     &2477.413     &2470.023     
    &2468.413     &2454.484     &2444.678     &2425.891  &2414.598\\     
    &2409.308     &2398.143     &2383.593     &2380.054  &2373.934     
    &2363.527     &2341.447     &2321.627  \\
   &2311.420     &2307.632     
    &2297.550     &2241.223     &2226.891     &2190.736  &2144.527 & \\    
\hline
\hline
\end{tabular}
\end{center}

\vspace{10mm}
\begin{center}
{\small Table V. Obtained parameters for 
infrared transition intensities of SiF$_4$ }
 
\vspace{3mm}
\begin{tabular}{ccccccc}
\hline
\hline
 &$\gamma_s$ & $\gamma_b$ & $\gamma_{sb}$ & $\eta_s$ & $\eta_b$ & SD\\
\hline
Fit A ~& ~39.276~ & ~21.642~ & ~61.850~ & ~3.578~ & ~3.281~ & ~1.745 \\
Fit B ~& ~37.472~ & ~21.441~ & ~61.944~ & ~3.527~ & ~3.285~ & ~2.015 \\
\hline
\hline
\end{tabular}
\end{center}

\end{document}